# RADAR: A TTP-based Extensible, Explainable, and Effective System for Network Traffic Analysis and Malware Detection

Yashovardhan Sharma
University of Oxford
United Kingdom
yashovardhan.sharma@cs.ox.ac.uk

Simon Birnbach
University of Oxford
United Kingdom
simon.birnbach@cs.ox.ac.uk

Ivan Martinovic
University of Oxford
United Kingdom
ivan.martinovic@cs.ox.ac.uk

## ABSTRACT
Network analysis and machine learning techniques have been widely applied for building malware detection systems. Though these systems attain impressive results, they often are (*i*) not extensible, being monolithic, well tuned for the specific task they have been designed for but very difficult to adapt and/or extend to other settings, and (*ii*) not interpretable, being black boxes whose inner complexity makes it impossible to link the result of detection with its root cause, making further analysis of threats a challenge. In this paper we present RADAR, an extensible and explainable system that exploits the popular TTP (Tactics, Techniques, and Procedures) ontology of adversary behaviour described in the industry-standard MITRE ATT&CK framework in order to unequivocally identify and classify malicious behaviour using network traffic. We evaluate RADAR on a very large dataset comprising of 2,286,907 malicious and benign samples, representing a total of 84,792,452 network flows. The experimental analysis confirms that the proposed methodology can be effectively exploited: RADAR's ability to detect malware is comparable to other state-of-the-art non-interpretable systems' capabilities. To the best of our knowledge, RADAR is the first TTP-based system for malware detection that uses machine learning while being extensible and explainable.

## CCS CONCEPTS

• **Security and privacy** → **Intrusion detection systems**; **Network security**; **Systems security**.

## KEYWORDS
Malware, Cyber Security, Explainable AI, Network Analysis

## 1 INTRODUCTION

Network analysis and machine learning (ML) techniques have been widely applied for building malware detection systems, see, *e.g.* [9, 16, 23, 28], for recent surveys covering the topic. The standard workflow in their development consists of gathering available data, then selecting and extracting a set of relevant features from it, building, training, and validating the ML model, before finally deploying into production. Though the resulting systems attain very positive results, they often are (*i*) monolithic, well tuned for the specific task they have been designed for, but requiring a complete retraining if not rewriting of the ML model in order to be adapted and/or extended to other attacks, and (*ii*) not explainable, since it is often not possible for humans to understand the reasons behind the model's predictions, or how to link the result of detection with its input features, especially when the input is in the form of low-level properties. Indeed, such systems are not designed taking into account the *extensibility* software design principle, based on the fact that not everything can be designed/foreseen in advance and thus calls for a modular design, or the principle of *explainability* and the related *interpretability* requirements, which impose that the inner mechanisms and the whole behaviour of the ML components have to be human understandable. Extensibility, explainability, and interpretability are particularly important properties to be satisfied by systems in the field of cyber security. Indeed, malware detection systems need to be continuously updated in order to deal with novel attacks, which, when detected, often need to be further analysed and inspected by domain experts (see also [3]). Despite the above motivations, all the systems for malware detection have not been designed to be extensible, and the vast majority of them are neither explainable nor interpretable.

In this paper we present RADAR, an explainable, interpretable, and extensible system for network traffic analysis and malware detection. Explainability and interpretability are obtained by using high-level, human understandable features of the network traffic, interpretable ML decision trees for malware detection, while extensibility is the result of the modular design of the malware detection engine which allows us to modify/extend one of its component without affecting the others. In particular, RADAR (*i*) exploits the popular TTP ontology of adversary behaviour described in the industry-standard MITRE ATT&CK framework to unequivocally identify and classify the potential malicious network flows in each traffic capture, (*ii*) for each TTP it utilises a dedicated decision tree to label each network flow as potentially malicious, and, (*iii*) combines the results obtained for the different TTPs in order to finally label each traffic capture as benign or malicious. As result, it is relatively easy to extend the system with new components for detecting newly discovered malicious TTPs and/or modify existing ones in order, *e.g.*, to broaden their applicability to variations of existing TTPs. Further, thanks to the combined use of TTPs and decision trees, RADAR is both an interpretable and explainable system, being able to output in human-understandable terms, (*i*) which are the high-level TTPs (and not just the low-level features) being used in the malicious detected traffic capture, and (*ii*) why the capture has been labelled as malicious. RADAR is the first system exploiting TTPs for automatic malware detection and is consequently the only system able to output which high-level TTPs are used in the detected malware's behaviour.

To evaluate RADAR's effectiveness, we created a very large dataset comprising of 2,286,907 malicious and benign samples, for a total of 84,792,452 network flows. The experimental analysis confirms that the proposed methodology can be effectively exploited: RADAR's ability to detect malware is comparable to other state-of-the-art non-extensible, non-interpretable systems' capabilities, as reported in the literature (see, *e.g.*, [18]). For instance, RADAR's performance as judged by an AUC score of 0.87 is similar to what has been reported in the very recent paper by Dyrmishi et al. [7].

Summing up, the key contributions of the paper are:

(1) we are the first to design an extensible, explainable, and interpretable system able to provide a malware classification according to a well-known, publicly available ontology of malware TTPs,
(2) we implemented RADAR, the first state-of-the-art system adhering to the above two requirements. RADAR is also the first system for automatic malware detection based on TTPs.
(3) we validate the ideas presented in this paper by evaluating RADAR over a very large real-world dataset of malware and benign samples, and thereby demonstrate its effectiveness.

The rest of the paper is structured as follows. Section 2 is dedicated to the presentation of the relevant related works. Section 3 provides a detailed description of RADAR design, also presenting some of the technological hypotheses and choices that guided its implementation. RADAR is then evaluated in Section 4 according to different metrics and with different settings. Section 5 contains a discussion on the results of the development and testing of RADAR. We end the paper with the conclusions and future work.

## 2 RELATED WORK

In existing literature, work related to utilising TTPs has so far focused on using TTPs to track the evolution of malware and describing emerging trends [5, 22], discovering inter-dependencies in a TTP chain [1], locating TTPs in the control flow graph describing the execution flow of a malware executable [8], automatically extracting TTPs from Cyber Threat Intelligence (CTI) reports [12, 17, 20], or detecting Living-Off-the-Land (LotL) malware techniques and APT attack campaigns [11, 15].

In the network domain, work has mostly centred around the extraction of MITRE ATT&CK TTPs from network traffic. For instance, McPhee [19] used the network IDS Zeek [24] to extract MITRE ATT&CK TTPs. Their paper discusses methods to identify several TTPs but provides only prototypical analysis scripts to test a few of these TTPs. BZAR [21] is another such example which has similar capabilities, as it enables the extraction of certain MITRE ATT&CK TTPs from Zeek log files. While RADAR also has a TTP Extraction phase, it goes significantly further than existing systems. For instance, just comparing this phase alone—McPhee's scripts identify 3 techniques, while BZAR can identify 12 techniques and 8 sub-techniques across 9 tactics. RADAR on the other hand utilises a selection of 9 techniques and 6 sub-techniques suited to NetFlow traffic, that altogether encompass 11 tactics from the ATT&CK framework. Compared to BZAR where a majority of techniques (13/20) cover just 3 tactics, RADAR is thus more suitable to detecting a wider variety of malicious activity. Additionally, unlike BZAR where most TTPs (18/20) rely primarily on DCE-RPC protocol messages to enable detection, RADAR does not depend upon any one protocol for its detection capabilities and hence can handle more versatile network traffic. Furthermore, while BZAR only allows for the configuration of the 'epoch' threshold used to set the period over which it analyses data to detect TTPs, RADAR supports a higher degree of configurability by virtue of:

(*i*) supporting thresholds for specific TTPs, (*ii*) allowing the selection of different detection policies for the overall system, and (*iii*) having policy-specific thresholds, which altogether enable a granular balance between the true-positive and the false-positive rate. Lastly, RADAR goes even further by using these TTPs to effectively classify malware based on NetFlows. While doing so, it provides an extensible design, explicitly separates malicious usage of TTPs from benign usage, and uses interpretable models to provide an explainable classification of maliciousness to an analyst using the system.

## 3 SYSTEM DESIGN

RADAR is a system for network traffic analysis and malware detection, designed to be both extensible and explainable. Extensibility is a well known software engineering and systems design principle that provides for future growth and adaptation. In the field of cyber security, and malware analysis in particular, extensibility is a necessary criterion to design systems that can adapt in the face of novel and constantly-evolving attacks. Explainability has recently emerged as an important requirement for ML and AI systems. As stated in [10], explainability is essential for users to effectively understand, trust, and manage powerful artificial intelligence applications in every application domain, including cyber security (see, e.g., [3, 4]). Explainability calls for producing high-level outputs which can be rooted back to the inputs, i.e., for exploiting ML interpretable models for the classification of malicious activities according to a well known ontology. With such goals, RADAR has been designed to

(1) be extensible, given its modular software architecture in which different TTPs are detected and classified by different ML engines, each of which can be easily extended and/or modified independently from the others, and
(2) provide high-level explanations, given its ability to classify malicious activity according to the TTP ontology provided by industry-standard MITRE ATT&CK framework and its use of interpretable decision trees for the classification.

RADAR's data pipeline is designed to be able to take arbitrary packet captures as input, and output:

(1) the relevant matching TTPs for that sample, and
(2) a prediction as to whether the sample is malicious.

This enables RADAR to be employed as a 'plug-and-play' technology wherein a packet capture can be sliced from a larger flow of network traffic and analysed. To further facilitate integrating RADAR into existing security infrastructures, it has been designed and implemented to also satisfy the following requirements:

(1) Be lightweight and as efficient as possible to enable the storage and possible retention of large amounts of data.
(2) Be content agnostic, *i.e.*, be based only on the analysis of the metadata, thus also working on encrypted data payloads and not violating possible privacy regulations that would otherwise limit the applicability of the system.
(3) Have the ability to analyse arbitrary network captures, *e.g.*, independent of traffic type, protocol, or network topology.

The components of RADAR include an Ingestion Processing System that constructs a graph database of malware samples, a

RADAR: A TTP-based Extensible, Explainable, and Effective System

| Field | Property |
|---|---|
| Flow Hash | The hash associated with each unique flow. |
| Sample Hash | The hash of the sample from which the flow is derived. |
| Unique ID | A unique ID associated with each flow. |
| Dataset | The dataset to which a flow belongs. |
| Source Port | The port from which the flow originates. |
| Destination Port | The port on which the flow connects to. |
| Start time, end time | Flow start or end time. |
| Duration | Flow duration in seconds. |
| Protocol | IP protocol identifier in decimal format. |
| Entropy | The Shannon Entropy for the flow payload. |
| Applabel | The application label, as identified by YAF. |
| Source IP, Dest. IP | Source and destination IPv4/IPv6 address. |
| Type, Code | ICMP type or code in decimal format. |
| Isn, Risn | Forward or reverse TCP sequence number. |
| Flags | 4 properties representing various TCP flags. |
| Tag, Rtag | 802.1q VLAN tag in forward/reverse direction of flow. |
| Pkt, Rpkt | No. of packets in forward/reverse direction of flow. |
| Oct, Roct | No. of bytes in forward/reverse direction of flow. |
| RTT | Round-trip time estimate in milliseconds. |
| End-reason | Reason for termination of flow. |

**Table 1: Flow properties.**

| Tactics | Techniques | Malicious | Benign | Class. |
|---|---|---|---|---|
| **Reconn.** | T1590 - Gather Victim Netw. Inf. | 4 | 4 | ✗ |
| **Cred.Access** | T1557.001 - Man-in-the-Middle | 7 | 0 | ✗ |
| **Discovery** | T1124 - System Time Discovery | 1,645,252 | 103,461 | ✓ |
| | T1135 - Network Share Discovery | 8224 | 41 | ✓ |
| **Lateral Movement** | T1021.001/4 - Remote Services(RDP/SSH) | 666 | 20 | ✗ |
| | T1550.003 - Use Alt. Auth. Material | 241 | 1 | ✗ |
| | T1563.001/2 - Rem.Serv.Sess.H.(RDP/SSH) | 3 | 0 | ✗ |
| | T1570 - Lateral Tool Transfer | 1 | 0 | ✗ |
| **Command and Control** | T1071 - Application Layer Protocol | 1997 | 15 | ✓ |
| | T1090 - Proxy | 965 | 2 | ✗ |
| | T1105 - Ingress Tool Transfer | 938 | 163 | ✓ |
| | T1571 - Non-Standard Port | 1,970,237 | 119,501 | ✓ |
| **Execution** | T1053 - Scheduled Task/Job | 10 | 5 | ✗ |

**Table 2: Supported MITRE ATT&CK TTPs.** The first column shows the tactical goals of the adversary. The second column shows the (sub-)techniques used to achieve them. The third and fourth columns indicate the absolute number of malicious and benign samples per TTP in our dataset. The last column indicates whether it is feasible to train a classifier for the given TTP.

TTP Detection Engine based on the MITRE ATT&CK framework, and a TTP-based Classification System. In the following subsections we shall cover each of the system components.

## 3.1 Data Ingestion and Graph Construction

The first stage in our data pipeline is the Ingestion System that takes as input network traffic captures (PCAPs) and builds a graph database of samples which preserves the most important characteristics of the traffic for the purposes of TTP detection. Among the various network data formats that we considered, we select NetFlows which rely solely on metadata, thus meeting the requirements of being content-agnostic and lightweight. In particular, we use a specific type of NetFlow, Yet Another Flowmeter (YAF) [13]. YAF generates bidirectional flows in IPFIX [6] standard, which is based on the Cisco NetFlow V9 export protocol. Our primary motivation for selecting YAF is that it is designed for high performance and scalability across large networks. By design, it also takes into consideration the balance between capturing information and maintaining privacy on the network, and takes a hybrid approach between typical NetFlows and PCAPs. It is also more feature-rich than the typical NetFlow or network log, since it contains support for functionality such as application labelling (*i.e.*, port independent protocol checking) and entropy analysis (to determine if traffic is encrypted or compressed). The complete list of features, including those extracted using YAF, are presented in Table 1.

Having selected and created the relevant features for our network traffic, we then map them into a graph whose nodes corresponds to the hosts as uniquely identified by their IP address, and whose edges correspond to a flow between the two hosts. This information is stored in a Neo4j graph database [1], which we query using the Cypher query language.

## 3.2 TTP Detection

The TTP Detection Engine analyses all flows in the graph database and executes detection rules on the graph entities and relationships to identify any matching MITRE ATT&CK technique. The first two columns of Table 2 show the complete list of TTPs that our system currently supports (notice that T1021 and T1563 have two sub-techniques each).

The set of rules utilised by our TTP Detection Engine has been selected being a good representative of the standard malicious activity carried out on networks, diverse enough in order to stress the ability of the system to detect different (also in terms of frequency) attacks. The TTPs we consider broadly falls into two categories:

(1) **Feature-based detection rules.** This type of detection relies on specific features of the flows described in Table 1.
(2) **Heuristic-based detection rules.** This type of detection relies on the analysis of multiple flows and uses heuristics based on their properties.

We demonstrate the utility of both types of rules and the specific logic required with the help of two examples.

*Example 3.1.* For the feature-based rules, we consider the MITRE ATT&CK technique T1571 (Non-Standard Port). This technique is part of the "Command and Control" (C&C) tactic and is applicable to the case when a host attempts to communicate with another host using standard network protocols but on a non-default port. Malware can attempt such a connection for several reasons—to communicate using a random port, evade firewall rules blocking certain default ports, or to find an open port for the purpose of establishing a C&C channel, among others. In order to detect this technique we rely on the features of the flow the two hosts use to communicate with each other. Specifically, we compare the network protocol in the flow with the port(s) used during communication and determine if these correspond to their defaults. In order to do this, we utilise YAF during the Ingestion Processing phase to determine the application layer protocol of every flow independent of the port used. This feature is extracted and stored in our graph as the flow property `applabel`. Our detection logic maintains a list of the 50 most popular application layer protocols and their corresponding default ports, and compares these to the features of `applabel` and `port` present in every flow. A mismatch between these features is a positive match for T1571.

---
[1]https://neo4j.com/product/neo4j-graph-database/

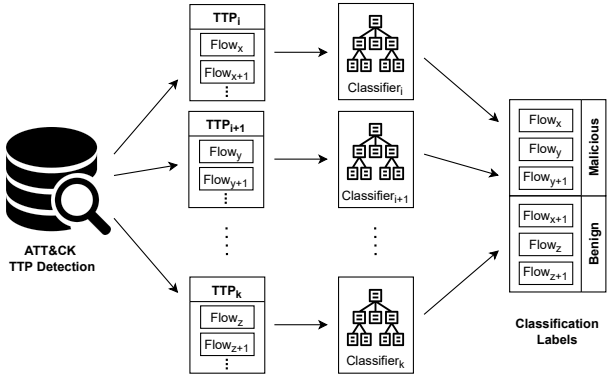

Figure 1: TTP-based Classification System Pipeline

*Example 3.2.* For the heuristic-based rules, we consider the MITRE ATT&CK technique T1071 (Application Layer Protocol). This technique is also part of the C&C tactic, and is applicable when an application layer protocol is used to hide malicious traffic. Specifically, we attempt to identify this technique by classifying flows as anomalous on the basis of their incoming and outgoing traffic relative to other flows. For each malware sample, we isolate its flows and establish a mean packet outflow to packet inflow ratio through their directed edges, and determine a unique baseline for each sample. Then by defining a threshold over which there is far more data egress than ingress as compared to the baseline, we identify flows and hosts that are potentially communicating via a C2 channel or exfiltrating data from the network. This threshold is designed to be dynamically adjusted based on the distribution of the network traffic and the characteristics of what is considered "normal" in a given network or environment.

At the end of this process, the system outputs a list of flows and their corresponding TTPs. This information is further aggregated by combining all the flows belonging to the same sample and then passed on to the TTP-based Classification system.

### 3.3 TTP-based Classification System

The TTP-based Classification System checks whether the flows in each sample exhibit malicious behaviour when compared to other flows matching the same TTP, and then flags a sample as malicious if "enough" flows exhibit malicious behaviour. The TTP-based Classification System works at deployment time following the steps in Figure 1.

As shown in the figure, the TTP-based Classification System takes as input: (*i*) the flows' properties (as listed in Table 1), and (*ii*) the TTPs detected in each flow, as returned by the TTP Detection Engine. Following a "divide-and-conquer" approach, it consists of one base classifier for each analysed TTP. This choice allows us to compare the behaviour of each flow with only those flows that match the same TTPs.

For each flow $f$, we create a vector $x$ containing all the features listed in Table 1. Then, $x$ is passed to all the base classifiers that have been trained on the flows that matched the same TTPs as $f$. Thus if, for example, $f$ matched $TTP_i$, $TTP_{i+1}$ and $TTP_k$ then $x$ will be passed just to the $i$th, $(i+1)$th and $k$th classifiers. Each of the selected classifiers then independently decides whether $f$ presents malicious behaviour when compared to the other flows flagged with the same TTP. Notice that in general any classifier can be used for this task. However, in order to preserve the transparency and explainability of the system, we opt to use decision trees [31]. Indeed, a decision tree is a transparent model whose predictions can be easily explained via Boolean logic rules. This is not true for most classifiers (*e.g.*, based on artificial neural networks) which are often completely opaque models whose predictions are difficult to interpret. As a result of this choice, it is possible to see exactly which network properties are used to make a classification decision, and this information can be quite useful to SOC analysts by helping them triage alerts in a more informed manner.

Once we have the prediction for each flow in the sample, we have to find a way to aggregate them and decide whether the sample is malicious or not. To this end, we devise three policies:

(1) P1: the first policy relies on $n_t$, the number of unique TTPs matched by the malicious flows in the sample. If $n_t \geq \theta_t$ then the sample is considered malicious. $\theta_t$ is a parameter that the user can tune.
(2) P2: the second policy relies on $n_f$, the number of malicious flows in the sample. If $n_f \geq \theta_f$ then the sample is considered malicious. $\theta_f$ is a parameter that the user can tune.
(3) P3: the third policy relies on $p$, the percentage of malicious flows in the sample. If $p \geq \theta_p$ then the sample is considered malicious. $\theta_p$ is a parameter that the user can tune.

The first policy, P1, is devised on the ground of the hypothesis that the more unique TTPs a sample maliciously uses, the more likely it is that the sample is engaged in malicious activity. However, this policy can become quite strict very quickly, as it is unlikely that the maliciousness of a sample will increase linearly with the number of TTPs after a certain point. We thus devise an alternate policy, P2, which is based on the number of malicious flows in a sample. The benefit of such a policy is that it does not rely solely on the presence of unique TTPs, and instead takes into consideration the fact that the overall repeated malicious usage of TTPs might be a good indicator of the sample's maliciousness. However, this policy's reliance on absolute values makes it susceptible to missing samples with lower numbers of malicious flows. Hence we devise our third policy, P3, which takes into account the percentage of malicious flows in the sample. It has similar benefits as P2 by virtue of utilising the repeated usage of TTPs and not relying solely on the presence of unique TTPs, however it uses the relative proportion of malicious flows in the overall sample to determine maliciousness. Notice that integrating other custom policies is quite straightforward, *e.g.*, assigning different weights to the TTPs and/or to the multiple uses of the same TTP. Thus, by designing and tuning policies for specific contexts, further performance gains can be extracted.

## 4 EVALUATION

We take a similar approach to many well-known systems in the literature when evaluating RADAR, wherein we test our detection capabilities against real-world malware datasets [24]. In doing so, we use an especially large dataset of network flows collected from malware samples for testing. We execute our samples within a Cuckoo sandbox on VirusTotal. All samples are allowed to freely create any network connections they wish, and the network traces of their actions are collected. Each sample is executed for two minutes within the Cuckoo sandbox, as research has shown this is sufficient for observing the majority of a sample's behaviour [14]. The resultant PCAPs corresponding to each sample are then given as input to RADAR in order to commence analysis. Table 3 shows the complete list of datasets, along with their properties, that we use to test our system. Ember is an open dataset of hashes of malicious Windows executables created as a benchmark for machine learning models. It contains 3 types of hashes: malicious, benign and unlabelled. For our purposes, we utilise both the malicious hashes (as Ember Malicious) and the benign hashes (as Ember Benign) in our dataset. MalRec is a small dataset created as a result of a running malware on a dynamic analysis platform, which was collected over a two-year period. MalShare is a community-driven open repository of malware samples, which features over a million hashes, from which the corresponding dataset is derived. Similarly, VirusShare is a repository that aims to help security researchers analyse selected strains of malware, and the source of our corresponding dataset. We select these datasets as they are representative of modern commodity malware and allow for the reproducibility of results. It is important to note that we do not utilise the entirety of each dataset in our experiments, since we filter out samples that do not exhibit any network activity. Overall, we analyse a total of 84,792,452 flows from 2,286,907 malicious and benign samples.

| Dataset | Samples | Flows |
| --- | --- | --- |
| Ember Malicious [2] | 435,741 | 26,567,527 |
| MalRec [27] | 37,763 | 12,273,502 |
| MalShare [25] | 1,268,923 | 32,310,907 |
| VirusShare [30] | 595,098 | 12,217,493 |
| Ember Benign [2] | 155,432 | 1,423,023 |
| Total (unique) | 2,286,907 | 84,792,452 |

Table 3: Overview of all the datasets. The first four datasets contain only malicious samples, while the last one contains only benign samples.

## 4.1 TTP Extraction

**Overview of Detected Techniques.** We deploy RADAR against the datasets shown in Table 3 to detect the TTPs our system supports. Table 2 shows the instances of the various MITRE ATT&CK TTPs that RADAR is able to detect in both our malicious and benign datasets and the total number of samples for which the corresponding TTP is detected. As seen in Table 2, we are able to detect nearly every type of MITRE ATT&CK TTP that RADAR supports in our dataset, though there is large variance in their frequency.

**Variance of Techniques Across Datasets.** We further explore the distribution of different techniques across all our datasets. Figure 2 shows the percentage of samples matching each MITRE ATT&CK TTP in every dataset. We find that certain TTPs have a high prevalence in all our datasets (*e.g.*, T1124 and T1571), while others occur with varying frequency depending on the specific dataset. These heterogeneous distributions are to be expected since each of these datasets is from different sources, and indicate that for the best performance, the system has to be tuned depending on the specific operational context. While the individual prevalence of certain techniques in our malware datasets shows the unique composition of the types of malware in each dataset, the frequency of occurrence of certain techniques in our benign dataset showcases some of the differences between the TTPs utilised by malicious and benign samples. For instance, T1550 (Use Alternate Authentication Material), T1557 (Man-in-the-Middle), and T1563 (Remote Service Session Hijacking) can be found in the MalShare and VirusShare datasets, while of these only T1550 is detected in Ember Malicious, while MalRec contains none of these TTPs. On the other hand for instance, the occurrence of T1090 (Proxy) is relatively infrequent in the benign dataset when compared to the malicious datasets. This would logically follow since benign samples are unlikely to utilise proxy services like Tor as frequently as malware samples.

**Malicious Usage of TTPs.** Our results highlight that techniques such as T1571 (Non-Standard port), T1124 (System Time Discovery) and T1135 (Network Share Discovery) are the most frequently occurring across all our malicious datasets. Other techniques used by malware in significant numbers include T1071 (Application Layer Protocol), T1021 (Remote Services), T1105 (Ingress Tool Transfer), and T1090 (Proxy). Whilst we discover evidence of other TTPs being used by malware, they are far less prevalent in their usage, but still indicate usage by a non-negligible amount of samples. The most commonly occurring TTP in our dataset is T1571 (Non-Standard Port). This shows a high prevalence of malware making use of ports that are not the default original source or destination port for a given application protocol. Whilst this technique has been written about in various papers and threat intelligence reports by antivirus firms (see, *e.g.*, [26, 29]), our pervasive detection of T1571 in all the different datasets affirms that this is a technique used by a large variety of different malware samples 'in-the-wild'.

**Benign Usage of TTPs.** We also observe a high number of TTPs being detected in our benign dataset. Specifically, T1571 (Non-Standard Port) and T1124 (System Time Discovery) are by far the most frequently occurring TTPs in our benign dataset. This is not surprising considering that benign software can be easily responsible for both these behaviours: (*i*) many applications can use non-standard ports to establish connections via commonly-used protocols (such as connecting via HTTPS on a port other than 443) and, (*ii*) several applications rely on having an accurate measure of time (say by contacting an Network Time Protocol server) to function correctly. This empirically highlights the challenge of using solely TTPs to determine the maliciousness of a given behaviour.

## 4.2 Classification performance

In this subsection, we evaluate how well the proposed system can distinguish between malicious and benign use of TTPs on the real-world dataset obtained by merging the datasets listed in Table 3.

**Experimental Setup.** *Dataset Creation per TTP*. Given such a dataset, we first need to assess whether enough flows/samples are matching a TTP to train the corresponding classifier. The results of this analysis are reported in the last column of Table 2. If we do not have enough flows/samples matching a TTP to train the corresponding classifier, we just classify each flow matching the TTP as malicious, and then we use the desired policy to conclude whether the sample is malicious or not.

For those TTPs for which training a classifier is feasible (*i.e.*, TTPs: T1071, T1105, T1124, T1135 and T1571) we then need: (*i*) a

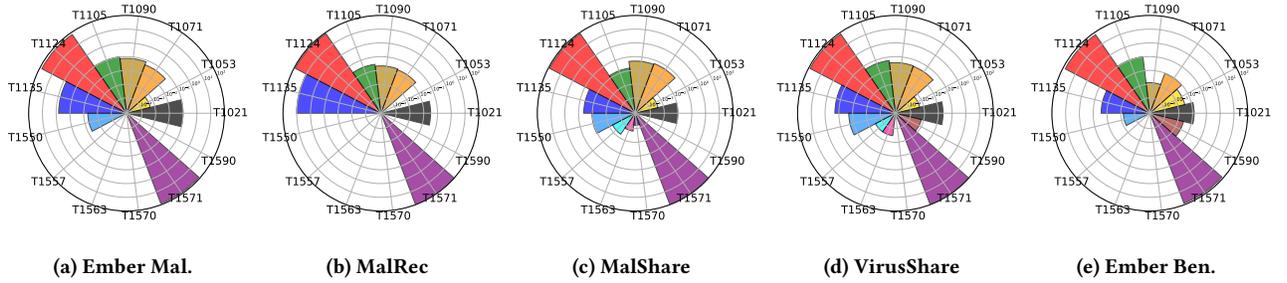

**(a) Ember Mal.**  **(b) MalRec**  **(c) MalShare**  **(d) VirusShare**  **(e) Ember Ben.**

**Figure 2: Percentage of samples matching various MITRE ATT&CK TTPs across each dataset, with a logarithmic Y-axis.**

"benign"/"malicious" label for each flow, and (ii) a training, validation and test set for each classifier. Regarding the labels, we simply assign a flow the label "malicious" (resp. "benign") if the sample to which it belongs is malicious (resp. benign). Regarding the creation of the datasets splits, we need to make sure that: (i) flows from the same sample do not belong to different splits, (ii) flows matching multiple TTPs belong to the same split for the different classifiers. To achieve the above we perform two steps. Firstly, we split the dataset obtained as the union of the datasets in Table 3 in train, validation and test sets, such that: the train set contains 60% of the samples, the validation set 10% of the samples, and the test set 30% of the samples. Secondly, for each flow $f$ in the training set and for each TTP $T$, we check whether $f$ matched $T$, and if that is the case, then $f$ is added to the training set of the classifier for the TTP $T$. The same procedure is then repeated for all the flows in the validation and test set. This ensures that the splits for each classifier retain the desired properties.

*Decision Tree Training.* Each decision tree is trained on the training set of the relevant TTP by maximising the Gini gain at each splitting step. In order to alleviate the imbalance problem across our malicious and benign datasets, we associate weights to each of the two classes that are inversely proportional to the class frequency in the training set. Further, for each decision tree associated to a TTP, we optimise two hyper-parameters: (i) the maximum depth of the decision tree, and (ii) the minimum number of samples in each leaf. For the maximum depth we try all the possible depths between 1 and 50, together with the option of not setting a maximum depth at all, while for the minimum number of samples we test values 1, 100 and 1,000. For each TTP, we then return the hyper-parameters that achieved the highest F1-score for the given TTP. While any other metric can be chosen, we pick F1-score in order to strike a balance between precision and recall (i.e., true positive rate). To ensure the reproducibility and robustness of our results, we initialise the training of each of our decision trees with five different random states, and then subsequently proceed with validation and testing for each state. The random state parameter controls the randomness of the decision tree. We average out the results from each random state for every decision tree and report only the averaged values going forth. We also compute 95% confidence intervals for each metric, however we do not plot these intervals as the range of the standard deviations is quite small. Indeed, the maximum standard deviation we register across all policies and thresholds is equal to $1.1 \times 10^{-4}$, thus indicating the stability of our models.

**Comparison of Different Aggregation Policies.** The first analysis we conduct is a comparison of the different aggregation policies described in Section 3.3. In order to have a complete understanding of the performance of each policy for each of the chosen thresholds, we evaluate the system according to 5 metrics: (i) accuracy, (ii) precision, (iii) true positive rate, (iv) false positive rate, and (v) F1-score. Each of these metric is necessary to fully understand the capabilities of the system and to tune the policy parameters as necessary. For example, choosing a higher threshold will probably minimise the false positive rate, however, it is equally important to understand the impact this will have on the true positive rate. In Figure 3/top we plot the performance of each of the policies according to the listed metrics while varying the decision thresholds. As we can see from Figure 3/top/left, P1 has good performance for low $\theta_t$ (i.e., $\theta_t = 1$ and $\theta_t = 2$), however, as expected, it soon becomes too strict and its true positive rate drops below 0.01 for $\theta_t = 3$. On the other hand, as we can see from Figure 3/top/middle and Figure 3/top/right, this does not happen for either P2 or P3. Further, we observe that all the policies start at almost the same values for our metrics (notice that P1 and P2 are equivalent for $\theta_t = \theta_f = 1$), but then different policies guarantee different levels of sensitivity when tuning their respective parameters, with P3 presenting the smoothest trend.

**Comparison with Exclusively TTP-based Approach.** The second analysis we conduct is to evaluate the need of the decision trees. To this end, we compare the results obtained by RADAR with Rule Based RADAR (RB-RADAR), which is RADAR without the TTP-based classification system. Exactly like RADAR, RB-RADAR is able to match each flow to the TTPs listed in Table 2. However, instead of using decision trees, RB-RADAR considers each flow matching at least one TTP as malicious. Then, in order to conclude whether a sample is malicious or not, it uses the same 3 policies—P1, P2, P3—as RADAR. For this analysis we use all the same metrics as before, and we plot the results in Figure 3/bottom.

Firstly, we notice that for $\theta_t = \theta_f = \theta_p = 1$, RB-RADAR achieves not only high accuracy, precision, true positive rate, and F1-score, but also a very high false positive rate (nearly 0.8). In contrast, for the same thresholds, RADAR obtains a false positive rate below 0.2 while keeping the value associated to the other metrics high. Secondly, we observe that when RB-RADAR achieves a false positive rate less than 0.2 (with P1 for $\theta_t = 3$, with P2 for $\theta_f = 5$, and with P3 for $\theta_p = 45$), this results in a drastic reduction in accuracy, true positive rate and F1-score, which all become less than 0.2 as well (the precision, similar to RADAR, is very high no matter the chosen threshold). This shows that using only TTPs for malware classification is an unsuitable approach, and using a two-step classification system based on decision trees, is a far better strategy.

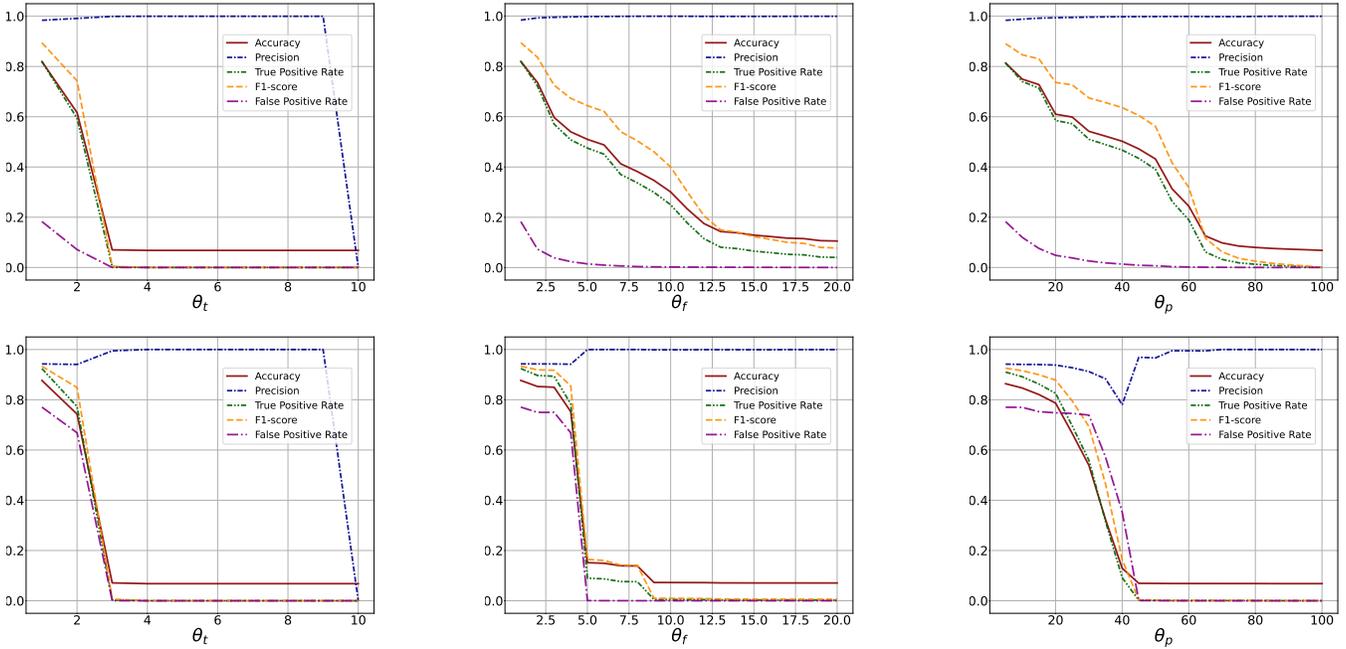

Figure 3: RADAR's (top figures) and RB-RADAR's (bottom figures) performances when using the 3 aggregation policies P1 (left), P2 (middle) and P3 (right) while varying the user defined threshold.

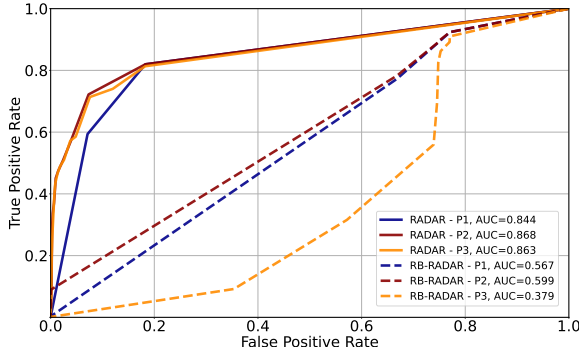

Figure 4: ROC Curves comparing RADAR and RB-RADAR.

Finally, in order to have a threshold independent comparison we plot the receiver operating characteristic (ROC) curve for each system and each of the three policies. We further compute the area under the curve (AUC) for each of these ROC curves. The results of this comparison are plotted in Figure 4. As we can see from the figure, while RADAR is able to achieve an AUC always greater than 0.8, the performance of RB-RADAR is close to random classification when using P1 and P2, and worse than random classification when using P3. This further highlights that using decision trees to classify malware on the basis of TTPs improves performance while maintaining a balance between true positive and false positive rate.

## 5 DISCUSSION

The development and testing of our system has interesting applications for malware analysis and the measurement of malware behaviour. In this section we discuss the results of our system development and testing.

**Prevalence of Techniques.** Based on our TTP extraction results, we find that only 8.62% of samples across all our datasets had no MITRE ATT&CK TTPs detected within them. While for just our malicious datasets, this number drops further to 7.56%. This suggests that TTPs are highly prevalent in malware and thus can be exploited for malware detection. The ATT&CK TTPs that we discovered to be prevalent within our datasets are also interesting. For instance, the use of non-standard ports to mask or blend-in malicious flows has been described by many different threat intelligence reports and papers in isolated incidents and small datasets. However our research concludes that this is a common technique across the vast majority of malware in all of our datasets.

**Variance of Techniques.** Our results show that there is variance among the techniques used within each dataset. The frequency of these techniques is different since the underlying nature of the commodity malware datasets differs from one another. We find that some of these TTPs are more common across datasets, but also that the distribution of these TTPs is not uniform. This variation can be used to evaluate the TTPs that are currently being employed by malware authors and concentrate resources for mitigation accordingly. At the same time, by studying the different usage of TTPs across benign and malicious samples, we gain a better understanding of the TTPs that are primarily used by only malware and hence are stronger indicators of malicious activity.

**Decision Trees and Performance.** Our results highlight the importance of exploiting decision trees in our ML pipeline. We find that a system relying solely on TTPs can have a false positive rate almost as high as 0.8, if tuned to maximise other metrics such as true positive rate, precision, accuracy, and F1-score. On the other hand, using a classification pipeline with decision trees reduces the false positive rate to less than 0.2 in the worst case, where each of the other metrics is maximised. As discussed earlier, any classifier can be used for the classification of the malicious activity in RADAR. However, when we performed some experiments using other popular ML models (in particular, Adaptive Boosting, Logistic Regression, Deep Neural Networks, Random Forest, EXtreme Gradient Boosting, and Support Vector Machines), we found that there was no significant variation in the performance, and indeed resulted in a less transparent and thus less explainable system.

## 6 CONCLUSIONS AND FUTURE WORK

In this paper we propose RADAR, a system that is capable of extracting MITRE ATT&CK TTPs from arbitrary network captures and using them to determine if the network traffic represents malicious behaviour. RADAR has been designed to be extensible and explainable. We demonstrate the effectiveness of RADAR by testing it against a dataset comprising of over 2 million samples, showing that RADAR is able to detect malware with an AUC score of 0.868. Our experimental results also highlight that TTPs occur frequently in both malware and benign samples. As a result, TTPs alone are insufficient to detect malicious activity and decision trees are fundamental in achieving these results. To the best of our knowledge, RADAR is the first system that is able to effectively and automatically separate malicious usage of TTPs from benign usage, while also using interpretable models to provide an explainable classification of maliciousness to an analyst using the system. Given the modularity of RADAR, other TTPs, machine learning models, and/or policies can easily be implemented and tuned for the specific intended use-case of the system. Indeed in future work, we plan to exploit these properties to further increase performance by (*i*) adding more TTPs, since this should correspond to a better profiling of each sample, and (*ii*) experimenting with other policies which are tuned for a specific use-case, e.g., by assigning weights to the TTPs and tuning them on the basis of the behaviours we want to detect. As per some preliminary experiments, we do not expect significant improvements in performance by using other machine learning (or even deep learning) models, which in any case come at the high price of sacrificing the interpretability of the system.